\documentclass[a4paper,11pt]{article}
\pdfoutput=1 

\usepackage{jheppub} 

\usepackage[T1]{fontenc} 

\usepackage{pdfsync}
\usepackage{mathptmx}
\usepackage[utf8]{inputenc}
\usepackage[T1]{fontenc}
\usepackage{slashed}
\usepackage{times}
\usepackage{amssymb,amsfonts,amsmath,amsthm}
\usepackage{dsfont,bbm}
\usepackage{dcolumn}
\usepackage{epsf}
\usepackage{graphicx}
\usepackage[caption=false]{subfig}
\usepackage{dsfont}
\usepackage{simplewick}
\usepackage{bm}
\usepackage{eucal}
\usepackage{slashed}
\usepackage[active]{srcltx}
\usepackage[usenames]{color}

\title{\boldmath Inverse Radon transforms: analytical and Tikhonov-like regularizations of inversion 
}

\author[a, b]{I.~V.~Anikin,}
\author[b, c]{Xurong~Chen}
\affiliation[a]{Bogoliubov Laboratory of Theoretical Physics JINR, 141980 Dubna, Russia}
\affiliation[b]{Institute of Modern Physics, Chinese Academy of Sciences, 730000 Lanzhou, P.R. China}
\affiliation[c]{Southern Center for Nuclear Science Theory (SCNT),
Institute of Modern Physics, Chinese Academy of Sciences, 516000 Huizhou, P.R. China}

\emailAdd{anikin@theor.jinr.ru}

\abstract{
We study the influence of analytical regularization used in the generalized function (distribution) space
to the Tikhonov regularization procedure utilized in the different versions of Moore-Penrose's inversion.
By introducing a new analytical term to the Tikhonov regularization
of Moore-Penrose's inversion procedure, we derive new optimization conditions that extend the
Tikhonov regularization framework and influence the fitting parameter.
This enhancement yields a more robust and accurate reconstruction of physical quantities, demonstrating its potential impact on various studies. We illustrate the significance of new term through schematic examples of physical applications,
highlighting its relevance to diverse fields.
Our findings provide a valuable tool for improving inversion methods and their applications in physics and beyond.
}

\begin{document}
\maketitle
\flushbottom

\section{Introduction}
\label{Intro}

As well-known, the mathematical basis of reconstructive (computerized) tomography is formed by
the direct and inverse Radon transforms \cite{Deans}. The reconstructive tomography gives us a possibility
to explore the internal structure of a given object without breaking it.
However, the inverse Radon transformation is more difficult mathematical task
compared to the direct Radon transformation
\footnote{For the Radon transforms, the inverse operators have different representations depending on the
space dimension, even or odd dimension.}:
\begin{eqnarray}
\label{Intro-1}
{\cal R}[f] = G \quad \Longrightarrow \quad f ={\cal R}^{-1} G
\end{eqnarray}
because we are forced to deal with the inverse operator that demands some suitable regularization.

Frequently, in the most practical cases, the inverse transforms suffer from the ill-posedness.
According Hadamard's criteria, some mathematical problems are called to be well-posed ones
provided they fulfil the following conditions: (a) existence ({\it i.e.} a solution of the problem exists)
(b) uniqueness ({\it i.e.} the given solution is unique) and
(c) stability ({\it i.e.} the given solution depends on the data only-continuously)
\cite{Bellmann, Kazakovtsev, Baguer, Kabri}. In opposite cases, we deal with the ill-posedness.

Moreover, in order to fully-reconstruct the internal distribution
we have to implement the infinite number of Radon transforms
\footnote{This statement is a consequence of the indeterminacy theorem: a given origin function $f$ with the compact support
is uniquely determined by any infinite set of its projections.}.
Of course, ``infinite'' is impossible from the practical point of view, therefore
the infinite number should be replaced by a reasonable number which is the other aspect
of difficulties related to the Radon transforms.

To invert the Radon transform, in both even and/or odd dimension of space,
there are the standard methods to do that (see for example \cite{Deans, GGV}).
Within the standard methods, in order to derive the inverse Radon transforms
we have to integrate with the symmetric angular measure:
\begin{eqnarray}
\label{Intro-2}
{\cal R}^{-1} \sim \int_{|\xi|=1} d\xi(\varphi) \, .... \, ,
\end{eqnarray}
where $\xi(\varphi)$ is defined by the corresponding angle $\varphi$ and $\varphi\in[0,\, 2\pi]$. Here, the explicit integrand
that is not important at the moment is omitted.
Notice that the integration measure of (\ref{Intro-2})  is mostly associated with the emission imaging where the radiation source is located inside
the scanned object. It cannot, however, be realized or might be realized with some difficulties in many cases of practices.

The most practically attractive kind of computerized tomography is provided by the transmission imaging where the radiation source is out of the scanned object.
In this case, as shown in \cite{Anikin:2019oes}, any well-localized original internal function (distribution), which describes the scanned object,
must lead to the angular restrictions on the integration measure of inverse Radon transforms.
In other words, the angular measure of integration for ${\cal R}^{-1}$ receives the angular constraints,
$\varphi\in[-\pi/2,\, \pi/2]$. As a result, we deal with the additional new term $f_A$ contributing the inverse Radon transforms that
is absent in the standard method.

Previously, in \cite{Anikin:2019oes}, the practical application of new term $f_A$ in QFT
has been demonstrated in the case where the transverse momenta of partons inside hadrons are
very essential. In the present paper, we show a role of new term $f_A$ in the Moore-Penrose inversion procedure
where the Tikhonov-like regularization has been utilized.

\section{The inverse Radon transformations and analytical regularization}
\label{Analytic}

In this section, we consider the analytical regularization of the inverse Radon transform which is inspired
by the method used for the singular generalized (distribution) functions.
For the sake of simplicity, we begin with the inverse Radon transforms \cite{Anikin:2019oes} in two-dimensional space.
As well-known, to derive the inverse Radon transform we first have to invert the Fourier transform given by
\begin{eqnarray}
\label{F-t-1-dir}
\mathcal{F}[f](\vec{\bf q})= \int_{-\infty}^{+\infty} d^2 \vec{\bf x} \, e^{-i\langle\vec{\bf q},\vec{\bf x}\rangle} \,f(\vec{\bf x}).
\end{eqnarray}
Generally speaking, both $\mathcal{F}[f](\vec{\bf q})$ and $f(\vec{\bf x})$ can be the complex functions,
{\it i.e.} $\{ \mathcal{F}[f](\vec{\bf q}),\, f(\vec{\bf x})\}\in \mathds{C}$.

It is worth to notice that the Fourier (direct or inverse) transform is always a good starting point for
many known transformations. Indeed, with the help of the corresponding replacements of variables,
the Fourier transform gives
the Laplace transforms (due to the replacement $i x = \omega$ in the Fourier transforms) or
the Borel transforms (due to the replacement $\omega=1/y$ in the Laplace transforms) and so on.

Thus, the inversion procedure yields
\begin{eqnarray}
\label{Inv-F-t-1}
f(\vec{\bf x}) = \int_{-\infty}^{+\infty} d^2 \vec{\bf q}\,  e^{+i\langle\vec{\bf q},\vec{\bf x}\rangle} \,\mathcal{F}[f](\vec{\bf q})
\Big|_{\vec{\bf q}=\lambda\vec{\bf n}_\varphi}=
\int_{0}^{+\infty} d\lambda \lambda \int_{0}^{2\pi} d\varphi\, e^{+i\lambda\langle \vec{\bf n}_\varphi, \vec{\bf x}\rangle}\,
\mathcal{F}[f](\lambda, \varphi)
\end{eqnarray}
where the polar coordinates have been used.
We now remind the Fourier slice theorem that merely states 
(in what follows the integral measures with brackets include the corresponding normalization factors
which are omitted unless it leads to misunderstandings)
\begin{eqnarray}
\label{F-t-3-dir}
\mathcal{F}[f](\lambda, \varphi) =\int_{-\infty}^{+\infty}(d\tau) \,e^{-i\lambda\tau}\,
\mathcal{R}[f](\tau,\varphi),
\end{eqnarray}
where
\begin{eqnarray}
\label{F-t-4-dir}
\mathcal{R}[f](\tau,\varphi)=
\int_{-\infty}^{+\infty} d^2 \vec{\bf x} \, f(\vec{\bf x})
\delta\left( \tau - \langle\vec{\bf n}_\varphi, \vec{\bf x}\rangle\right)
\end{eqnarray}
defines the direct Radon transformation of $f(\vec{\bf x})$.
Using (\ref{F-t-3-dir}), after simple algebra (\ref{Inv-F-t-1}) can be presented in a form of
\begin{eqnarray}
\label{Inv-F-t-2}
f(\vec{\bf x}) =
 \int_{0}^{2\pi} d\varphi \,
\int_{-\infty}^{+\infty} (d\eta)\,
\mathcal{R}[f](\eta + \langle \vec{\bf n}_\varphi, \vec{\bf x}\rangle, \varphi)
\int_{0}^{+\infty} d\lambda \lambda \, e^{-i\lambda\eta}.
\end{eqnarray}
From this, we can readily observe that the integration over $\lambda$ suffers from the
singularity at the infinite integral limit.
In other words, this integral representation demands some regularization.
It is not something unexpected in the generalized function theory where
the regularization is introduced by $\eta\to\eta - i\epsilon$. This is the so-called {\it analytical regularization}.
Given that, for the $\lambda$-integration, one gets
\begin{eqnarray}
\label{int-lam}
i\frac{\partial}{\partial\eta} \int_{0}^{+\infty} d\lambda \, e^{-i\lambda(\eta-i\epsilon)}\equiv 2\pi\frac{\partial}{\partial\eta} \delta_-(\eta)
=
-\frac{\mathcal{P}}{\eta^2} + i\pi \frac{\partial}{\partial\eta}\delta(\eta),
\end{eqnarray}
where $\lambda$ has been traded for the derivative over $\eta$ acting on the exponential function.

We finally derive the regularized inverse Radon transform which can be written as
\begin{eqnarray}
\label{Inv-F-t-3}
f(\vec{\bf x}) \Rightarrow  f_\varepsilon(\vec{\bf x})=f_S(\vec{\bf x}) + f_A(\vec{\bf x}),
\end{eqnarray}
where
\begin{eqnarray}
\label{Inv-F-t-3-a}
&&f_S(\vec{\bf x})=
-\int_{-\infty}^{+\infty} (d\eta)\, \frac{\mathcal{P}}{\eta^2}\,
\int_{0}^{2\pi} d\varphi \, \mathcal{R}[f](\eta + \langle \vec{\bf n}_\varphi, \vec{\bf x}\rangle, \varphi)
\end{eqnarray}
is the standard contribution, while
\begin{eqnarray}
\label{Inv-F-t-3-b}
f_A(\vec{\bf x})=
-i\pi \int_{-\infty}^{+\infty} (d\eta) \,  \delta(\eta)\, \int_{0}^{2\pi} d\varphi
\Big[ \frac{\partial}{\partial\eta}\mathcal{R}[f](\eta + \langle \vec{\bf n}_\varphi, \vec{\bf x}\rangle, \varphi) \Big]
\end{eqnarray}
corresponds to the new term found in \cite{Anikin:2019oes} {\it provided that the angular integration is limited by the
interval $[-\pi/2, \, \pi/2]$} instead of the full interval $[0,\, 2\pi]$. That is, it can be proven that
\begin{eqnarray}
\label{Inv-F-t-3-bb}
\boxed{
f_A(\vec{\bf x}) \not= 0 \hspace{0.3cm} \text{iff}\quad d\varphi \Rightarrow d\varphi \,\Theta\big( \varphi\in[-\pi/2, \, \pi/2]\big)
\hspace{0.2cm}
\text{in (\ref{Inv-F-t-3-b})},
}
\end{eqnarray}
where $\Theta(...)$ is a generalized version of the standard $\theta$-function, {\it i.e.} 
$\Theta=1$ if the condition represented by the argument takes place, otherwise $\Theta=0$.

It is worth to notice that the regularizations based on the analytical continuation methods 
lead naturally to the appearance of the imaginary part (see for example \cite{Gelfand:1964}).
It can be readily understood from the consideration 
of the generalized function given by $1/x$ with a singularity at $x=0$.
As well-known, the regularized version of this function can be written as 
\begin{eqnarray}
\label{reg-over-x}
\frac{1}{x+i0} \equiv \lim_{\varepsilon\to 0}\, \frac{1}{x+i\varepsilon}=
\frac{{\cal P}} { x} - i\, \pi \delta(x).
\end{eqnarray}
We stress that (\ref{int-lam}) is merely the integral representation of the mentioned function.
Moreover, working within the analytical continuation method of the theory of generalized functions, 
it is possible to demonstrate that the imaginary part 
given by $-\pi \delta(x)$ is nothing but the corresponding residue of $F(x)=\varphi(x)/x$, where $\varphi(x)$ 
belongs to the finite function space and forms the functional of generalized functions.

If $\mathcal{R}[f] \in \mathbb{C}$ and $f(\vec{\bf x})\in \Re{\rm e}$, 
the origin function $f(\vec{\bf x})$ should be directly replaced by the approximated function 
$f_\varepsilon(\vec{\bf x})$. However, in the case of $\mathcal{R}[f] \in \Re{\rm e}$ and $f(\vec{\bf x})\in \Re{\rm e}$,
the necessary condition of the nullification of $f_A(\vec{\bf x})$ leads to the specific properties of $\mathcal{R}[f]$.
Both these cases have been considered in detail in \cite{Anikin:2019oes}.
In the present paper, we study the different role of 
the imaginary additional term $f_A(\vec{\bf x})$ that does contribute to the new condition
extending the Tikhonov regularization, see below.

As explained in \cite{Anikin:2019oes},
the restriction of (\ref{Inv-F-t-3-bb}) on the angular integration can be traced from the fact that the compact origin function $f$ has
led to the angular restriction for its Fourier image. In its turn, the angular dependences of Fourier and Radon transforms are equivalent
to each other, see (\ref{F-t-3-dir}).

In (\ref{Inv-F-t-3-b}) and (\ref{Inv-F-t-3-bb}), these equations actually gives the integral representation of $f_A$,
the derivative over $\eta$ can be dragged
to $\mathcal{R}[f]$ without the surface terms. It follows from the theorem asserting that the direct Radon transform of the function with
the localized (or restricted) support is also a compact function \cite{Deans}.

\section{The operator form of direct and analytically-regularized inverse 
Radon transforms}

For our further purposes, let us rewrite the direct and inverse Radon transforms in a form of operators which 
act on the suitable (Hilbert) space.
At the first, we introduce the operator of direct Radon transforms as (c.f. (\ref{F-t-4-dir}))
\begin{eqnarray}
\label{An-Reg-Op-1}
G=\mathcal{ R} [f]\quad \text{with} \quad
\mathcal{ R} [...]=\int_{-\infty}^{+\infty} d^2 \vec{\bf x} \,
\delta\left( \tau - \langle\vec{\bf n}_\varphi, \vec{\bf x}\rangle\right) [....].
\end{eqnarray}
We here emphasize the singular character of 
the nullified radial Radon parameter $\tau$. Indeed, it can be understood as the degeneration of the nullified radial parameter
in the polar system of coordinates. 
To illustrate this point, let us consider the example of $\mathbb{R}^2$ where two vectors $X$ and $Y$ emanate from the same starting point $O$.
In the frame of polar (or spherical) system, we have
$X(r_X, \phi_X)\equiv (r_X \cos\phi_X, r_X\sin\phi_X)$ and the similar representation can be written for $Y$.
If the radius vectors of both  $X$ and $Y$ are not equal to zero, even if they are infinitesimal ones, it is possible to distinguish these two vectors.
However, if $r_X=r_Y=0$, the starting point $O$ loses somehow information on the differences of vectors $X$ and $Y$ due to
the condition $O=(|\vec{\bf 0}|\,\cos\phi_X, |\vec{\bf 0}|\,\sin\phi_X)=(|\vec{\bf 0}|\,\cos\phi_Y, |\vec{\bf 0}|\,\sin\phi_Y)$.
It results in the degeneration of the starting point $O$ (or $\tau=0$) in the Radon transforms (see also \cite{Anikin:2021oht})
which requires the suitable regularization.
The regularized operator $\mathcal{ R}_\varepsilon$ corresponding to the direct Radon 
transform has been introduced below with the help of additional arguments, see (\ref{F-t-4-dir-reg-1}).

Then, the inverse Radon transforms can be presented in the form of  
(c.f. (\ref{Inv-F-t-3})-(\ref{Inv-F-t-3-bb}))
\begin{eqnarray}
\label{irt-1}
f=\mathcal{ R}^{-1}[G] ,
\end{eqnarray}
while the (analytically) regularized inverse Radon transforms take the operator form as
\begin{eqnarray}
\label{reg-irt-1}
&&f_\epsilon=\mathcal{ R}^{-1}_\epsilon[G]\quad \text{with} \quad
\nonumber\\
&&
\mathcal{ R}^{-1}_\epsilon[...]=
-\int_{-\infty}^{+\infty} (d\eta)\, \frac{\mathcal{P}}{\eta^2}\,
\int_{-\pi/2}^{\pi/2} d\varphi \, [...]
-i\pi \int_{-\infty}^{+\infty} (d\eta) \,  \delta(\eta)\, \int_{-\pi/2}^{\pi/2} d\varphi \,  [...].
\end{eqnarray}
In (\ref{An-Reg-Op-1})-(\ref{reg-irt-1}) the argument dependences are not shown explicitly.
The regularized representation of (\ref{reg-irt-1}) plays an important role in the formulation 
of the functional condition which extents the Tikhonov regularization.

\section{The dual Radon transforms and analytical regularization}

In this section, we present the arguments which underly the necessary regularization even for the 
direct Radon transforms.  For this purpose, 
we introduce the dual Radon transforms defined as
\begin{eqnarray}
\label{Dual-R-1}
\mathcal{R}^\ast[G]\equiv \mathcal{R}^\ast[G](\vec{\bf x}) =
\int_{-\infty}^{+\infty} d\tau \int_{0}^{2\pi} d\varphi \,G(\tau, \varphi)
\delta\left( \tau - \langle\vec{\bf n}_\varphi,\vec{\bf x}\rangle\right)
=\int_{0}^{2\pi} d\varphi \,G(\langle\vec{\bf n}_\varphi,\vec{\bf x}\rangle, \varphi),
\end{eqnarray}
where $G$ implies, generally speaking, an arbitrary function of two variables but, in our consideration, 
we assume that $G(\tau, \varphi)$ is being $\mathcal{R}[f](\tau, \varphi)$ giving the 
dual Radon transform of the Radon transform.
So, 
making used the replacement: $G(\tau, \varphi) \Rightarrow\mathcal{R}[f](\tau, \varphi)$, we write down the following
combination
\begin{eqnarray}
\label{Dual-R-2}
\mathcal{R}^\ast \mathcal{R}[f](\vec{\bf x}) =
\int_{0}^{2\pi} d\varphi \,\mathcal{R}[f](\langle\vec{\bf n}_\varphi,\vec{\bf x}\rangle, \varphi).
\end{eqnarray}
In order to see the other realization of analytical regularization applied to the Radon transform,
we now consider the inverse Fourier transform of $1/|\vec{\bf q}|$. We have
\begin{eqnarray}
\label{An-F-1}
{\cal F}^{-1}\Big[ \frac{1}{|\vec{\bf q}|} \Big]\Big|_{\vec{\bf q}=\lambda\vec{\bf n}_\varphi}
= \int_{0}^{2\pi}d\varphi \int_{0}^{+\infty} d\lambda \, e^{+i \lambda \langle\vec{\bf n}_\varphi,\vec{\bf x}\rangle}.
\end{eqnarray}
As in (\ref{Inv-F-t-2}), the integration over $\lambda$ requires the analytical regularization which finally leads to
\begin{eqnarray}
\label{An-F-2}
{\cal F}^{-1}_\epsilon\Big[ \frac{1}{\lambda} \Big]
= \int_{0}^{2\pi}d\varphi \, \frac{1}{\langle\vec{\bf n}_\varphi,\vec{\bf x}\rangle + i\epsilon}=
\int_{0}^{2\pi}d\varphi \int_{-\infty}^{\infty} d\tau \, \frac{1}{\tau+ i\epsilon}
\delta(\tau - \langle\vec{\bf n}_\varphi,\vec{\bf x}\rangle)\equiv {\mathcal R}^*\Big[ \frac{1}{\tau+ i\epsilon} \Big],
\end{eqnarray}
where the prescription for the radial Radon parameter has been also dictated by the necessity of 
the analytical regularization appled for the corresponding integral representation.
Hence, it hints the regularization for the direct Radon transforms as well, see below.

For ${\mathcal R}^\ast {\mathcal R}$-operator, it is now instructive to study the set of eigenvalues and eigenfunctions.
For the sake of simplicity, we again begin with the corresponding Fourier images of this operator,
we derive that (here, the argument dependence has been shown explicitly)
\begin{eqnarray}
\label{An-F-3}
{\cal F}\Big[ {\mathcal R}^\ast {\mathcal R}[G]\Big](\lambda \vec{\bf n}_\varphi)
=\frac{1}{\lambda} {\cal F}[G](\lambda \vec{\bf n}_\varphi) .
\end{eqnarray}
If we assume that $G(\vec{\bf x})=e^{+i \langle \vec{\bf Q},\vec{\bf x}\rangle}$ (this is merely a plane wave function),
then one can easily obtain that
\begin{eqnarray}
\label{An-F-4}
{\cal F}\Big[ {\mathcal R}^\ast {\mathcal R}[e^{+i \langle \vec{\bf Q},\vec{\bf x}\rangle} ]\Big](\vec{\bf q})
=\frac{1}{|\vec{\bf Q}|} \delta^{(2)}\big(  \vec{\bf q} - \vec{\bf Q}\big) .
\end{eqnarray}
or
\begin{eqnarray}
\label{An-F-5}
{\mathcal R}^\ast {\mathcal R}[e^{+i \langle \vec{\bf Q},\vec{\bf x}\rangle} ]
=\frac{1}{|\vec{\bf Q}|} e^{+i \langle \vec{\bf Q},\vec{\bf x}\rangle} .
\end{eqnarray}
From (\ref{An-F-5}) one can conclude that $1/|\vec{\bf Q}|$ is an eigenvalue of operator ${\mathcal R}^\ast {\mathcal R}$ and
the plane wave is its eigenfunction. 
Hence, it means that, first, the zero value of $|\vec{\bf Q}|$  should be regularized and, second, 
the singular value of ${\mathcal R}^{-1}$ is going to infinity if $|\vec{\bf Q}|\to \infty$,
{\it i.e.} (see Appendix~\ref{LSM-SVD:App:A} for the definition of singular values $\sigma_A$ of matrix/operator $A$)
\begin{eqnarray}
\label{inv-R-sv-1}
\sigma_{{\mathcal R}^{-1}}=\sqrt{|\vec{\bf Q}|}\to \infty,
\end{eqnarray}
requiring, as a result, the other regularization as well.
It is worth to note that since the inverse Radon operator is an unbounded, it leads also to
the amplification of noises, see Appendix~\ref{ill-pose-noise}.

To conclude this section, we take into account (\ref{An-F-2}) together with (\ref{inv-R-sv-1}) and infer
that in order to get the bounded support for the inverse Radon transforms
we have to make the definite regularization as $\tau \to \tau+i\epsilon$ applied to the Radon parameter. 
That is,
we deal with the following regularized Radon transforms
\begin{eqnarray}
\label{F-t-4-dir-reg-1}
\mathcal{R}_\epsilon[f](\tau,\varphi)\equiv \mathcal{R}[f](\tau + i\epsilon,\varphi)=
\int_{-\infty}^{+\infty} d^2 \vec{\bf x} \, f(\vec{\bf x})
\delta\left( \tau +i\epsilon - \langle\vec{\bf n}_\varphi, \vec{\bf x}\rangle\right) = G_\epsilon,
\end{eqnarray}
where we suppose that the integration variable can be analytically continued up to the complex values. 
The extension of the delta function, considered as the corresponding singular functional, 
on the complex argument can be found in  \cite{Gelfand:1964}
\footnote{
Alternatively, in the function space,
 the delta function of the argument with a complex prescription can be
also understood in the frame of sequential approach \cite{Anikin:2023kdh} where 
$\lim_{\varepsilon\to 0} \delta\left( x- A - i\varepsilon \right) \Rightarrow
\delta\left( i\, 0 \right) = \lim_{\varepsilon\to 0} \, \lim_{x \to A} \,1/(x- A - i\varepsilon )$. 
}.

Thus, 
we have demonstrated that the unbounded support of the Radon transforms is nothing but the singularity (degeneration) of the case where the radial Radon parameter goes to zero.
All these types of singularities demand the special regularizations which are proposed in the paper.

\section{The extended Tikhonov regularization of inversion}
\label{TR-pr}

Building on the analytical regularization of the inverse Radon operator presented in the preceding section,
we now explore alternative approaches to address the challenges of inverting singular operators.
While our regularization is inspired by the theory of generalized functions,
the other methods have been developed to deal with this problem.
One of such approaches is Tikhonov's regularization \cite{Kazakovtsev, Baguer, Kabri} which implements Moore-Penrose's inversion procedure
\cite{MP}
 through a different realization.
 This indirect inversion method offers a distinct perspective on the problem, underscoring the need for
 a comprehensive understanding of the various strategies which are available for addressing the singular operator inversion.

 \subsection{The Moore-Penrose (generalized) inversion: a short description}
 \label{subsec:MP}
 
As well-known from the matrix theory, the Moore-Penrose (generalized) inversion \cite{MP} allows to find a solution 
of 
\begin{eqnarray}
\label{MP-1}
{\rm A}\, x = y  
\end{eqnarray}
in the matrix representation even if the matrix ${\rm A}\in \mathbb{C}_n^{m x n}$ is a singular matrix, {\it i.e. }
$Ker\, ({\rm A})\not= \{0\}$  (or $det \,{\rm A}=0$).
Indeed, if we build the matrix ${\rm A}^{+} \, {\rm A}$ 
\footnote{Here, ${\rm A}^+$ implies the Hermitian conjugated matrix.}, 
which is not now the singular matrix of order $n$,
we can go over to the solution of ${\rm A}^{+} \, {\rm A} \, x = {\rm A}^{+}  y$  with the help of 
the least-squared approximation, see Appendix~\ref{LSM-SVD:App:A}.
The matrix $X=({\rm A}^{+} \, {\rm A})^{-1} {\rm A}^{+} $ is called the Moore-Penrose (generalized) inversion, ${\rm A}^\dagger$,
of the matrix ${\rm A}$ if it satisfies the Penrose conditions: (a) ${\rm A} X {\rm A}={\rm A}$, (b) $X {\rm A} X = X$,
(c) $({\rm A} X)^+ = {\rm A} X$ and (d) $(X {\rm A} )^+ = X {\rm A}$.

 \subsection{The standard Tikhonov regularization}

In the above-mentioned operator equation, see (\ref{MP-1}), the both sides of equation are known in principle.
In the practical applications, the {\it r.h.s.} of (\ref{MP-1}) can be only approximately known.
In this case, the inversion procedure should be supplemented by the Tikhonov regularization. 

To demonstrate the method in the context of our case, we consider the operator representation of the direct Radon transforms,
see (\ref{An-Reg-Op-1}). Let us now suppose that
the system given by (\ref{An-Reg-Op-1}) is not compatible one (this is rather a typical situation for the ill-posed inversion),
{\it i.e.} instead of (\ref{An-Reg-Op-1}) we have
\begin{eqnarray}
\label{An-Reg-Op-1-2}
G\approx {\cal R} [f].
\end{eqnarray}
According the normal equation theorem, see Appendix~\ref{LSM-SVD:App:A},
one can conclude that the best approximation for the solution of this system, first, can be found by minimization of
the norm
\begin{eqnarray}
\label{BApp-1}
\min_f \big\{ \Vert {\cal R} [f] - G \Vert \big\}
\end{eqnarray}
and, then, it is given by a solution of the system (if ${\cal R}^\ast \sim A^T$ and ${\cal R}\sim A$, see Appendix~\ref{LSM-SVD:App:A})
\begin{eqnarray}
\label{An-Reg-Op-1-3}
{\cal R}^\ast\, {\cal R} [f] = {\cal R}^\ast [G] \quad \Longrightarrow \quad f = \left[ {\cal R}^\ast\, {\cal R} \right]^{-1}{\cal R}^\ast[G].
\end{eqnarray}

It is important to emphasize that in contrast to the typical case considered in subsection~\ref{subsec:MP} 
the operator $[{\cal R}^\ast\, {\cal R}]^{-1}$ demands some additional regularization for
the well-definitness (boundedness) of solution $f$, see (\ref{An-F-5}). Fortunately, the standard Tikhonov regularization,
which improves originally the approximation to the exact solution, can naively regularized the mentioned singularity too. 
Indeed, we can introduce the regularized operator as
\begin{eqnarray}
\label{An-Reg-Op-1-4}
{\cal R}^\ast\, {\cal R} + \Lambda \,{\cal I} \quad \Longrightarrow \quad
f_{reg.} = \left[ {\cal R}^\ast\, {\cal R} + \Lambda \,{\cal I} \right]^{-1} {\cal R}^\ast[G],
\end{eqnarray}
where $\Lambda$ is a Lagrange factor, see below, and ${\cal I}$ is a unit operator.
This kind of regularizations implies that the norm of (\ref{BApp-1}), that should be minimized, has to be modified too.
Thus, the optimitization procedure should be applied for the following modified norm
(in the literature, it is usually called as Tikhonov's regularization):
\begin{eqnarray}
\label{BApp-2}
\min_f \big\{ \Vert {\cal R} [f] - G \Vert^2  + \Lambda \, \Vert f \Vert^2 \big\},
\end{eqnarray}
where $\Lambda$ should be chosen for the best approximation.

The similar condition of (\ref{BApp-2}) is wide-used for the different sorts of operators in (\ref{MP-1})
in the practical computations. As a rule, the only ill-posedness  of ${\rm A}^{-1}$ is associated with the breaking 
of continuity regarding some external parameters. 
However, the direct and inverse Radon transforms, which are given by the 
integral operators, stand aside due to the fact that the integral representations
are not well-defined themselves. In this connection,
the analytical regularization inspired by the theory of generated functions 
can be applied.

 \subsection{The extended Tikhonov regularization}

We are now in a position to focus on the generalization of Tikhonov's regularization.
On the ``physical'' level of rigor, 
we shortly remind the main items of approach where the approximated solution should be found.
We suppose that the exact {\it r.h.s.} of (\ref{An-Reg-Op-1-2}) exists, {\it i.e.}
\begin{eqnarray}
\label{Exact-Eq}
{\cal R} [f]=G_0.
\end{eqnarray}
As mentioned, in the practical cases, instead of $G_0$ we often deal with approximated values $G_{\delta_i}$, see (\ref{An-Reg-Op-1-2}).
We require that not only the variations between $G_0$ and $G_{\delta_i}$ are small but the variations   
inside the corresponding sets of approximated values are small enough too, {\it i.e.} 
\begin{eqnarray}
\label{mes-1}
\Vert G_{\delta_i} - G_{0} \Vert \leq \epsilon, \quad 
\Vert G_{\delta_1} - G_{\delta_2} \Vert \leq \epsilon, \quad \Vert f_{\delta_1} - f_{\delta_2} \Vert \leq \varepsilon.
\end{eqnarray}
Notice that the elements $f_\delta$ should minimize some external additional functional $\Omega[f]$ which has been chosen 
in a such way to fulfil the condition as $\Vert f_\delta \Vert^2 = \epsilon^2$.
The latter condition can be included in the minimizing function, see (\ref{BApp-2}), through the Lagrange factor $\Lambda$, see below.

We stress that in the standard Tikhonov approach the minimization procedure has been implemented 
in the spaces $U$, which is formed by $G_\delta$,
and $F$, which is formed by $f_\delta$, separately while the regularizing operator $\tilde R$ giving by 
\begin{eqnarray}
\label{Reg-Op-1}
f_\delta = \tilde R[G_\delta]
\end{eqnarray} 
is supposed to exist but it has been never used explicitly.

In the present study, we contrarily derive the regularizing (or regularized) inverse operator  ${\cal R}^{-1}_\epsilon$ (see (\ref{reg-irt-1}))
and, then, we insist the condition $\Vert \, {\cal R}^{-1}_\epsilon[G] \, \Vert^2 = \epsilon^2$ through the Lagrange multiplier method.
In this case, the Lagrange function reads
\begin{eqnarray}
\label{Lag-1}
{\cal L}_\Lambda = \Vert \, {\cal R} [f] - G \, \Vert^2 + \Lambda \, \Big(\Vert \, {\cal R}^{-1}_\epsilon[G] \, \Vert^2 - \epsilon^2\Big).
\end{eqnarray}   

Further, we explore the Lagrange function ${\cal L}_\Lambda$ on the conditional extremum regarding all variables to 
obtain the following condition 
\begin{eqnarray}
\label{BApp-3}
\boxed{
\min_f \big\{ \Vert {\cal R} [f] - G \Vert^2  + \Lambda \, \Vert \, {\cal R}^{-1}_\epsilon[G] \, \Vert^2 \big\}
}
\end{eqnarray}
which is called the extended Tikhonov regularization.

Alternatively, the optimization procedure can be applied to
\begin{eqnarray}
\label{BApp-4}
\min_f \big\{ \Vert {\cal R} [f] - G_\epsilon \Vert^2  + \Lambda \, \Vert f \Vert^2 \big\}.
\end{eqnarray}

The key findings of the present paper are encapsulated in the conditions presented by (\ref{BApp-3}) and (\ref{BApp-4}).
A close examination of these conditions reveals that the additional term 
stemmed from the analytically regularized inversion operator
plays a crucial role in the Tikhonov-type regularization. This term has a significant impact on the fitting parameter $\Lambda$.

The extension of Tikhonov's regularization defined by (\ref{BApp-3}) can be applied to the
iterative algorithms used for the reconstruction from projections. It is very close to the inversion problems considered in
Section~\ref{TR-pr}.
In this case, the direct Radon operator $\mathcal{R}$ (and its the regularized version $\mathcal{R}_\epsilon$)
has to be presented in the matrix form. To do that, let us rewrite the integral representation of direct Radon transforms as
\begin{eqnarray}
\label{R-tr-MF-1}
\mathcal{R} [f](\tau, \varphi)=\int_L d\sigma \, f\big( \tau \,\vec{\bf n}(\varphi) + \sigma \,\vec{\bf n}_\perp(\varphi) \big),
\end{eqnarray}
where $\vec{\bf n}(\varphi)=(\cos\varphi, \sin\varphi)$ and $\vec{\bf n}_\perp(\varphi)=(-\sin\varphi, \cos\varphi)$ 
and the integration is going along the line parametrized in the argument.

Now, we are going over to the digital picture of the origin function $f$. In other words, the function $f$ has been represented by its
digitized version (in pixels). Following \cite{Deans}, this kind of discrete approximations leads to the matrix forms given by (c.f. (\ref{R-tr-MF-1}))
\begin{eqnarray}
\label{R-tr-MF-2}
\mathcal{R}_{i,\,j} [f] \equiv
\mathcal{R} [f](\tau_i, \varphi_j)=\sum_{M=1}^n \sum_{K=1}^n  \Delta \sigma_{M\, K} \, f_{M\,K\,;\, i,\, j}\, ,
\end{eqnarray}
where summation goes over the pixel squares presented in the matrix form and  $\tau_i\, ,\varphi_j$ are temporally fixed but they finally vary
within the pixel picture. We also assume that $1< i < n$ and $1 < j < n^\prime$.
Notice that, alternatively, (\ref{R-tr-MF-2}) can be written in the matrix representation as
\footnote{The detail explanation of the transition between (\ref{R-tr-MF-2}) and (\ref{R-tr-MF-3}) has been presented in \cite{Deans}.}
\begin{eqnarray}
\label{R-tr-MF-3}
{\bf R} = \hat {\bf F} \,{\bf S},
\end{eqnarray}
where ${\bf R}$ is a column matrix with elements marked by $n\, n^\prime$, $ \hat {\bf F}$ defines the $n\,n^\prime \times n^2$ matrix
related to $f_{M\,K\,;\, i,\, j}$ and ${\bf S}$ is a column matrix with $n^2$ number of elements and associates with $ \Delta \sigma_{M\, K}$.
Given (\ref{R-tr-MF-3}), one can apply the extended Tikhonov regularization, see (\ref{BApp-3}), to reconstruct the (digital) structure
of the origin function $f$.

\section{The potential applications}
\label{Phys-Apps}

In this section, we explore the potential physical applications of the newly derived terms.
Notably, the direct Radon transform has been extensively utilized in the medical imaging,
in particular, in emission and transmission reconstructed tomography.
The logarithmic ratio of the input and output intensities, obtained after an X-ray or proton beam passes through the scanned object,
is equivalent to the direct Radon transform of the energy-fixed damping function
$f_{d}(x; \rho, Z)\equiv f_{d}(x)$
\footnote{Here, the fixed energy $E$ is omitted in the arguments; $\rho$ and $Z$ imply the material density and the atomic number, respectively.}, {\it i.e.}
\begin{eqnarray}
\label{Log-Inten-1}
- \log \Big[ \frac{I}{I_0}\Big] =  \int_{L(\tau,\varphi)} d\sigma \, f_d\big( x(\sigma); \rho, Z\big) \equiv {\cal R}[f_d](\tau,\varphi),
\end{eqnarray}
where $L(\tau, \varphi)$ implies the line parametrization with the radial and angular parameters $\tau$ and $\varphi$,
respectively.
This representation gives the so-called linear realization where the energy $E$ of beam is fixed.
The {\it l.h.s.} of (\ref{Log-Inten-1}) relates to the observable value measured by apparatus.

To circumvent the degeneracy point ($\tau=0$) of the Radon transform, we employ a suitable analytical regularization of the direct Radon transform,
${\cal R}[f_d] \to {\cal R}_\epsilon[f_d]$ as described by (\ref{F-t-4-dir-reg-1}).
This regularization gives an additional contribution to (\ref{Log-Inten-1}) due to the imaginary part of ${\cal R}_\epsilon[f_d]$.

The case of fixed X-ray energy is usually not a very good for the practical applications.
As a rule, the energy has some distribution $w(E)$ in the known interval.
It means that we practically deal with the following representation
\begin{eqnarray}
\label{Log-Inten-2}
 \frac{I}{I_0} = \int d\mu(E, w)\, \exp\Big\{ - \int_{L(\tau,\varphi)}  d\sigma \, f_d\big( x(\sigma), E)\Big\} \Rightarrow
 \int d\mu(E, w)  \, e^{ - {\cal R}_\epsilon[f_d\big( x(\sigma), E)]},
\end{eqnarray}
where the integration measure includes the energy averaging and  the corresponding weight function $w$.
This is already the non-linear realization due to the integration measure is presented.

A striking resemblance is observed between the representation of $ I / I_0$ given by (\ref{Log-Inten-2}) and
the case of essential transverse momenta in the corresponding parton
distribution functions
considered in QFT, see \cite{Anikin:2019oes}, where the imaginary part of ${\cal R}_\epsilon^{-1}$ (or ${\cal R}_\epsilon$)
generates an extremely important contribution.
Indeed, the most natural correspondence can be established by mapping the line position
$x(\sigma)$ to the parton longitudinal fraction $y$ and the X-ray (or beam) energy $E$ to the
parton transverse momenta $k_\perp$.
This correspondence enables a clearer understanding of the other sources of imaginary parts and highlights the potential for further connections between these seemingly disparate frameworks.

\section{Conclusions}
\label{Cons}

In this paper, we have demonstrated the significant role of the new term $f_A$  (or $f_\epsilon$) in the Moore-Penrose inversion procedure
which has been regularized using the Tikhonov-like approach. We have introduced the novel conditions, see (\ref{BApp-3}) and (\ref{BApp-4}),
on the optimization procedure
extending the scope of the Tikhonov regularization
and influencing the fitting parameter $\Lambda$ in the inversion process.
Our results have the far-reaching implications for various applications
where the additional term in $f_\epsilon$ (or in $G_\epsilon$), see (\ref{reg-irt-1}), arising from the analytical regularization,
can play a crucial role in different studies.

\section*{Acknowledgements}
We thank G.~I.~Andreev, Sh.~Z.~Sharipov, D.~Strozik-Kotlorz, L.~Szymanowski and Jian-Hui~Zhang
for useful and illuminating discussions.
Also, I.V.A. thanks his colleagues  from
the Chinese University of Hong Kong for a very warm hospitality.
The work has been supported in part by PIFI 2024PVA0110
Program.

\appendix
\renewcommand{\theequation}{\Alph{section}.\arabic{equation}}
\section*{Appendix}

\section{The manifestation of ill-posedness due to the noise }
\label{ill-pose-noise}

Let us consider the mathematical convolution of two functions written as 
$g=f_1 \,\otimes\, f_2 $ where the function $f_2$ is given by $f_2(x)=e^{-x^2/2}$ and its Fourier image is 
${\cal F}[f_2](\omega)=e^{-\omega^2/2}$. In a theory, we can recover the function $f_1$ from the function $g$ by the following 
inversion:
\begin{eqnarray}
\label{inv-1}
{\cal F}[f_1](\omega) = {\cal F}[g](\omega)\, e^{\omega^2/2},
\end{eqnarray}
where $e^{\omega^2/2}$ is nothing but ${\cal F}^{-1}[f_2](\omega)$.
However, in a reality, we may deal with the noise effect which can be expressed as
$g=f_1 \,\otimes\, f_2 + z$. Therefore, we have 
\begin{eqnarray}
\label{inv-2}
{\cal F}[f_1](\omega) = {\cal F}[g](\omega)\, e^{\omega^2/2} - {\cal F}[z] \,e^{\omega^2/2},
\end{eqnarray}
where the exponential factor in the second term ensures that we are grossly inaccurate if the noise $z$ presents.

\section{Least squares optimization and singular value decomposition}
\label{LSM-SVD:App:A}

In this appendix, for the pedagogical reason we remind the basic material of linear algebra that is needed for our study,
see for example \cite{Bellmann}.
Let us begin with an operator $A: H_1\to H_2$ where $H_i$ can be assumed to be the Hilbert space.
We want to find  $x\in H_1$ that satisfies the equation $Ax=y$. As well-known, a solution $x$ of this equation can be found
uniquely if and only if $Ker(A)={0}$ where $Ker(A): \{ x\in H_1 \,| \,Ax=0 \}$.

Now let $A$ be an $m\times n$ matrix with $m>n$ and $Rank(A)=n$. The least squares approximation (LSA) of the system $Ax\approx b$ is a vector $x$ that minimizes the norm given by $\Vert\, Ax-b\,\Vert$ where the norm is defined as $\Vert\, Y \,\Vert_p=(\sum_{i=1}^n |\, Y\, |^p)^{1/p}$ for $p\in\mathbb{Z}$.
An alternative definition of LSA can be formulated as follows: LSA of $Ax\approx b$ is a solution of the system given by
$A^T A x = A^T b$. This is {\it a theorem on the normal equation (or the normal equation theorem)}.

One of the fundamental method of matrix representation widely used in machine learning and data science is
the singular value decomposition (SVD). This decomposition allows to reduce the dimensionality and the noise influence
together with the realization of data vision.
To explain the main idea of SVD, we consider again
the above-mentioned matrix $A$. Then, we construct the combinations  $A\, A^T$ and $A^T\, A$ which are now
symmetric matrices and, therefore, both matrices can be orthogonally diagonalized, {\it i.e.}
$A \,A^T =U \,D_1 \,U^T$ and $A^T\, A = V\, D_2\, V^T$.
These decompositions give the singular value decomposition in a form of $A=U \Sigma V^T$
where $\Sigma$ is a diagonal $m\times n$ matrix containing the singular values which are, by definition,
equal to the square roots of the eigenvalues of $D_1$ and $D_2$.

Notice that all eigenvalues $\lambda$ of $A\, A^T$ and $A^T\, A$ (these matrices have the same set of eigenvalues)
are positive ones, {\it i.e.} $\lambda\ge 0$.
Moreover, if $p$ is an eigenvector of $A\, A^T$ with $|p|=\langle p,p \rangle=1$,
then $q=A^T p/\sigma$ is an eigenvector of $A^T\, A$ where
$\sigma=\sqrt{\lambda}$ defines the set of singular values of $A$.
One can easily see that it takes place the converse relation given by
$p=A \,q/\sigma$.


\end{document}